\documentstyle[aps,pra,epsf,multicol,amssymb,pstricks,latexsym,graphicx]{revtex}

\newcommand{\swapbit}[1]{\rput#1{$\times$}}

\newcommand{\Comment}[1]

\def\ket#1{| #1\rangle}

\def\S{{\cal{S}}}

\def\qed{\qquad{\bf \rm QED.}}

\def\R{\hbox{\rm I \kern-5pt R}}

\begin{document}

\title{A Comparison of Quantum Oracles} \author{Elham
Kashefi$^{*\dagger}$, Adrian Kent$^{\S}$, Vlatko Vedral$^{*}$ and
Konrad Banaszek$^{*\dagger}$} \address{$^{*}$Optics Section, The
Blackett Laboratory, Imperial College, London SW7 2BZ, England \\
$^{\dagger}$ Centre for Quantum Computation, Clarendon Laboratory,
University of Oxford, Parks Road, Oxford OX1 3PU, England \\ $^{\S}$
Hewlett-Packard Laboratories, Filton Road, Stoke Gifford, Bristol BS34
8QZ, England}
\date{\today}
\maketitle

\begin{abstract}
A standard quantum oracle $S_f$ for a general function $f: Z_N
 \rightarrow Z_N $ is defined to act on two input states and return
 two outputs, with inputs $\ket{i}$ and $\ket{j}$ ($i,j \in Z_N $)
 returning outputs $\ket{i}$ and $\ket{j \oplus f(i)}$. However, if
 $f$ is known to be a one-to-one function, a simpler oracle, $M_f$,
 which returns $\ket{f(i)}$ given $\ket{i}$, can also be defined. We
 consider the relative strengths of these oracles.  We define a
 simple promise problem which minimal quantum oracles can solve
 exponentially faster than classical oracles, via an algorithm which
 cannot be naively adapted to standard quantum oracles.  We show
 that $S_f$ can be constructed by invoking $M_f$ and $(M_f )^{-1}$
 once each, while $\Theta(\sqrt{N})$ invocations of $S_f$
 and/or $(S_f )^{-1}$ are required to construct $M_f$.
\end{abstract}

\draft

\begin{multicols}{2}
Recent years have witnessed an explosion of interest in quantum
computation, as it becomes clearer that quantum algorithms are more
efficient than any known classical algorithm for a variety of
tasks.\cite{Deutsch85,Shor94,Grover96,BBBV97}. One important way of
comparing the efficiencies is by analysing {\it query complexity},
which measures the number of invocations of an ``oracle'' --- which
may be a standard circuit implementing a useful sub-routine, a
physical device, or a purely theoretical construct --- needed to
complete a task. A number of general results show the limitations and
advantages of quantum computers using the query complexity models
\cite{BBCMW98,vanDam98,Cleve99}.

In this paper we compare the query complexity analysis of quantum
algorithms given two different ways of representing a permutation in
terms of a black box quantum oracle. We begin with a short discussion
of graph isomorphism problems, which motivates the rest of the paper.

Suppose we are given two graphs, $G_1 = (V_1 , E_1 )$ and $G_2 = (V_2
, E_2)$, represented as sets of vertices and edges in some standard
notation. The graph isomorphism (GI) problem is to determine whether
$G_1$ and $G_2$ are isomorphic: that is, whether there is a bijection
$f: V_1 \rightarrow V_2$ such that $( f(u) , f(v) ) \in E_2$ if and
only if $(u,v) \in E_1$. (We assume $| V_1| = | V_2 |$, else the
problem is trivial.)  GI is a problem which is NP but not known to be
NP-complete for classical computers, and for which no polynomial time
quantum algorithm is currently known.

We are interested in a restricted version (NAGI) of GI, in which it is
given that $G_1$ and $G_2$ are non-automorphic: i.e., they have no
non-trivial automorphisms. So far as we are aware, no polynomial time
classical or quantum algorithms are known for NAGI either. The
following observations suggest a possible line of attack in the
quantum case.

First, for any non-automorphic graph $G = (V,E)$, we can define a
unitary map $M_G$ that takes permutations $\rho$ of $V$ as inputs and
outputs the permuted graph $\rho(G) = (\rho(V), \rho(E))$, with some
standard ordering (e.g. alphabetical) of the vertices and edges, in
some standard computational basis representations. That is, writing $|
V | = N$, for any $\rho \in S_N$, $M_G$ maps $\ket{\rho}$ to
$\ket{\rho(G)}$. Consider a pair $(G_1 , G_2 )$ of non-automorphic
graphs. Given circuits implementing $M_{G_1}$, $M_{G_2}$, we could
input copies of the state $ \frac{1}{\sqrt{N!}} \sum_{\rho \in S_N}
\ket{\rho}$ to each circuit, and compare the outputs $\ket{\psi_i} =
\frac{1}{\sqrt{N!}}
\sum_{\rho \in S_N} \ket{ \rho (G_i ) }$. Now, if the graphs are
isomorphic, these outputs are equal; if not, they are
orthogonal. These two cases can be distinguished with arbitrarily high
confidence in polynomial time (see below), so this would solve the
problem.

Unfortunately, our algorithm for NAGI requires constructing circuits
for the $M_{G_i}$, which could be at least as hard as solving the
original problem. On the other hand, it is easy to devise a circuit,
$S_G$, which takes two inputs, $\ket{\rho}$ and a blank set of states
$\ket{0}$, and outputs $\ket{\rho}$ and $\ket{\rho (G) }$. Since $S_G$
and $M_G$ implement apparently similar tasks, one might hope to find a
way of constructing $M_G$ from a network involving a small number of
copies of $S_G$. Such a construction would solve NAGI. Alternatively,
one might hope to prove such a construction is impossible, and so
definitively close off this particular line of attack.

Thus motivated, we translate this into an abstract problem in query
complexity.

Consider the following oracles, defined for a general function $f: \{0,1\}^m
\rightarrow \{0,1\}^n$:
\begin{itemize}
\itemsep0pt \parsep0pt
\item  the {\it standard} oracle,  $ S_f : |x\rangle |b\rangle
\rightarrow |x\rangle |b \oplus f(x)\rangle$.
\item the {\it Fourier phase} oracle, $P_f : |x\rangle |b\rangle
\rightarrow e^{2\pi i f(x) b / 2^n }|x\rangle |b\rangle$.
\end{itemize}
Here $x$ and $b$ are strings of $m$ and $n$ bits respectively, represented
as numbers modulo $M=2^m$ and $N = 2^n$, $| x
\rangle$ and $| b \rangle$ are the corresponding computational basis
states, and $\oplus$ is addition modulo $2^n$.

Note that the oracles $P_f$ and
$S_f$ are equivalent, in the sense that each can be constructed by an
$f$-independent quantum
circuit containing just one copy of the other, and also
equivalent to their inverses.   To see this, define the quantum
Fourier transform operation $F$ and the parity reflection $R=F^2$ by
$$
F : \ket{j} \rightarrow \frac{1}{ \sqrt{N}}
\sum_{k=0}^{N - 1} \exp ( 2 \pi i j k / N )
\ket{k} \, , \qquad
R: \ket{j} \rightarrow \ket{-j} \, .
$$
Then we have
\begin{eqnarray*}
&(& I \otimes F ) \circ S_f \circ (I \otimes F^{-1} ) = P_f \, , \\
&(& I \otimes F^{-1} ) \circ P_f \circ (I \otimes F) = S_f \, , \\
&(& I \otimes R ) \circ S_f \circ (I \otimes R ) = (S_f )^{-1} \, , \\
& (& I \otimes R ) \circ P_f \circ (I \otimes R ) = (P_f )^{-1} \, .
\end{eqnarray*}

For the rest of the paper we take $m=n$ and suppose we know $f$ is a
permutation on the set $\{ 0,1 \}^n$. There is then a simpler
invertible quantum map associated to $f$:
\begin{itemize}
\itemsep0pt \parsep0pt
\item the {\it  minimal} oracle: $ M_f : |x\rangle \rightarrow |f(x)\rangle$.
\end{itemize}

We can model NAGI, and illustrate the different behaviour of
standard and minimal oracles, by a promise problem. Suppose we are
given two permutations, $\alpha$ and $\beta$, of $Z_N$, and a subset
$S$ of $Z_N$, and are promised that the images $\alpha(S)$ and
$\beta(S)$ are either identical or disjoint.  The problem is to
determine which. (This problem has been considered in a different
context by Buhrman et al \cite{BCWW01}.)

We represent
elements $x \in Z_N$ by computational basis states of $n$ qubits in
the standard way, and write $|S \rangle = \sum_{x \in S} \ket{x}$.

Figure $1$ gives a quantum network with minimal oracles that
identifies disjoint images with probability at least $1/2$.

\begin{figure}
\psset{unit=0.6cm}
\begin{pspicture}(0,8)(-1,0)
%\psgrid(10,10)
\psline(0.5,1)(10,1)
\psline(0.5,2)(9.5,2)
\rput(7,2.6){$\vdots$}
\rput(0.85,2.6){$\vdots$}
\psline(0.5,3)(9.5,3)
\psline(0.5,3.5)(9.5,3.5)
\psline(0.5,4)(9.5,4)
\rput(.2,3.05){$\displaystyle\left\{\rule{0pt}{0.8cm}\right.$}
\rput(-.6,2.9){$\displaystyle |S \rangle$}
\psline(.5,5)(9.5,5)
\rput(7,5.6){$\vdots$}
\rput(0.85,5.6){$\vdots$}
\psline(.5,6)(9.5,6)
\psline(.5,6.5)(9.5,6.5)
\psline(.5,7)(9.5,7)
\rput(.2,6.05){$\displaystyle\left\{\rule{0pt}{0.8cm}\right.$}
\rput(-.6,5.9){$\displaystyle |S \rangle$}
\rput(0,1){$\displaystyle |1 \rangle$}
\psline(4,1)(4,7)
\psline(5,1)(5,6.5)
\psline(6,1)(6,6)
\psline(8,1)(8,5)
\psframe[fillstyle=solid,fillcolor=white](1.2,1.8)(2.6,4.2)
\rput(1.9,3){$\scriptstyle O_{\beta}$}
\psframe[fillstyle=solid,fillcolor=white](1.2,4.8)(2.6,7.2)
\rput(1.9,6){$\scriptstyle O_{\alpha}$}
\psframe[fillstyle=solid,fillcolor=white](1.5,1.4)(2.3,0.6)
\rput(1.9,1){$\scriptstyle H$}
\pscircle(4,1){.1}
\pscircle(5,1){.1}
\pscircle(6,1){.1}
\pscircle(8,1){.1}
\psframe[fillstyle=solid,fillcolor=white](8.6,1.4)(9.4,0.6)
\rput(9,1){$\scriptstyle H$}
\swapbit{(4,7)}
\swapbit{(4,4)}
\swapbit{(5,6.5)}
\swapbit{(5,3.5)}
\swapbit{(6,6)}
\swapbit{(6,3)}
\swapbit{(8,5)}
\swapbit{(8,2)}
\pswedge(10,1){.4}{270}{90}
\end{pspicture}
\caption{A quantum circuit for the permutation promise
problem. $O_{\alpha}$ and $O_{\beta}$ are minimal oracles for
computing the permutations $\alpha$ and $\beta$ respectively,
$|S\rangle$ is the superposition of all the basis states, $H$ is the
Hadamard transformation, and all the other gates are conditional swap
gates, where circles signify control bits.}
\end{figure}
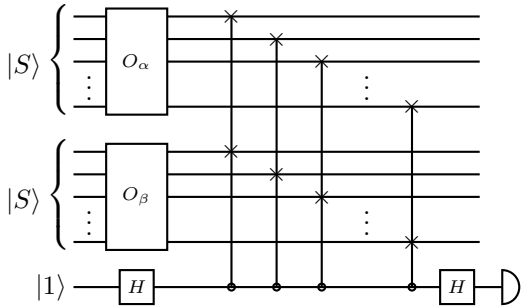

Let $A=\{
 \alpha(x)|x \in S \}$ and $B=\{ \beta(x)| x \in S \}$.  One query to
the oracles $M_{\alpha}$ and $M_{\beta}$ creates the (unnormalised) states
$| A \rangle$ and $| B \rangle$
respectively.  The state before applying the controlled gates is:
\begin{eqnarray*}
| A \rangle | B \rangle
\otimes (|0\rangle - |1\rangle )
\end{eqnarray*}
After controlled swap gates, the state becomes:
$$
| A \rangle | B \rangle |0\rangle -
| B \rangle | A \rangle |1\rangle  \, .
$$
The final Hadamard gate on the ancilla qubit gives:
$$
( | A \rangle | B \rangle  - | B \rangle | A \rangle ) |0\rangle
+ ( | A \rangle | B \rangle  + | B \rangle | A \rangle )  |1\rangle
$$

A $|0\rangle$ outcome shows unambiguously that the images are disjoint. A
$|1\rangle$ outcome is generated with probability $1$ if the images are
identical, and with probability $1/2$ if the images are disjoint.
Repeating the computation $K$ times allows one to exponentially improve
the confidence of the result. If after $K$ trials we get $|0\rangle$ at
least once, we know for certain that $\alpha(S) \neq \beta(S)$. When all
the $K$ outcomes were $|1\rangle$, the conclusion that
$\alpha(S)=\beta(S)$ has the conditional probability $p_K = \frac{1}{2^K}$
of having been erroneously generated by disjoint input images. Note that
$p_K$ is independent of the problem size and decreases exponentially with
the number of repetitions.

Clearly, a naive adaptation of the algorithm to standard oracles does
not work.  Replacing $M_{\alpha}$ and $M_{\beta}$ by $S_{\alpha}$ and
$S_{\beta}$, and replacing the inputs by $\ket{S} \otimes \ket{0}$,
results in output states which are orthogonal if the images are
disjoint, but also in general very nearly orthogonal if the images are
identical.  Applying a symmetric projection as above thus almost
always fails to distinguish the cases. To the best of our knowledge a
non-trivial lower bound for this problem using the $S_f$ is not known
(however, see \cite{Aaronson01}).

This example suggests that minimal oracles may be rather more powerful
than standard oracles. To establish a more precise version of this
hypothesis, we examine how good each oracle is at simulating the
other. One way round turns out to be simple.  We can construct $S_f$
from $M_f$ and $(M_f)^{-1} = M_{f^{-1}}$ as follows:
$$
S_f = ( M_{f^{-1}} \otimes I ) \circ A \circ ( M_f \otimes I ) \,
$$
where $\circ$ represents the composition of operations (or the concatenation of
networks) and the
modulo $N$ adder $A$ is defined by $A : \ket{a} \otimes \ket{b}
\rightarrow \ket{a} \otimes \ket{a \oplus b }$.

Suppose that we are given $M_f$ in the form of a specified
complicated quantum circuit.  We may be completely unable
to simplify the circuit or deduce a simpler form of $f$ from it.
However, by reversing the circuit gate by gate, we can construct
a circuit for $(M_f )^{-1}$.  Hence, by the above construction,
we can produce a circuit for $S_f$, using one copy and one
reversed copy of the circuit for $M_f$.

This way of looking at oracles can be formalised into
the {\it circuit model}, in which the query
complexity of an algorithm involving an
oracle $O_f$ associated to a function $f$ is the number of copies
of $O_f$ and/or $O^{-1}_f$ required to implement the algorithm in a
circuit that, apart from the oracles, is independent of $f$.
In the circuit model, a standard oracle can easily be simulated
given a minimal oracle.  Ignoring constant factors, we say that the minimal
oracle is at least as strong as the standard oracle.

It should be stressed that, while the circuit model has a natural
justification, there are
other interesting oracle models, to which our arguments will not
apply.  For example, if we think of the oracle $M_f$ as a black box supplied
by a third party, then we should not assume that $(M_f)^{-1}$ can easily be
constructed from $M_f$, as we know no way of efficiently reversing the
operation of an unknown physical evolution.

Remaining within the circuit model, we now show that
$M_f$ and $S_f$ are not (even up to constant factors) equivalent.
In fact, simulating $M_f$ requires exponentially many uses of $S_f$.

First, consider the standard oracle $S_{f^{-1}}$ which maps a basis
state $|y\rangle |b \rangle$ to $|y\rangle |b \oplus f^{-1}
(y)\rangle$.  Since $S_{f^{-1}} : |y\rangle |0 \rangle \rightarrow
|y\rangle | f^{-1}(y)\rangle$, simulating it allows us to solve the
search problem of identifying $|f^{-1}(y)\rangle$ from a database of
$N$ elements.  It is known that, using Grover's search algorithm, one
can simulate $S_{f^{-1}}$ with $O(\sqrt{N})$ invocations of $S_f$
\cite{BHT98,BHMT00}. In the following we explain one possible way of
doing that.

Prepare the state $|y\rangle|0\rangle|0\rangle|0\rangle$, where the
first three registers consist of $n$ qubits and the last register is a
single qubit. Apply Hadamard transformations on the second register to
get $|\phi_1 \rangle = |y\rangle\sum_{x \in
Z_N}|x\rangle|0\rangle|0\rangle \mbox{.}$ Invoking $S_f$ on the second
and third registers now gives $$ |y\rangle( \sum_{x \in
Z_N}|x\rangle|f(x)\rangle ) |0\rangle \mbox{.}$$ Using CNOT gates,
compare the first and third registers and put the result in the
fourth, obtaining
$$ \Big(|y\rangle\sum_{x \in Z_N , x \ne
f^{-1}(y)}|x\rangle|f(x)\rangle|0\rangle\Big)+
\Big(|y\rangle|f^{-1}(y)\rangle|y\rangle|1\rangle\Big) \mbox{.}$$
Now
apply $( S_f )^{-1}$ on the second and third registers, obtaining
$$
\Big(|y\rangle\sum_{x \in Z_N , x \ne
  f^{-1}(y)}|x\rangle|0\rangle|0\rangle \Big) +
\Big(|y\rangle|f^{-1}(y)\rangle|0\rangle|1\rangle \Big) \mbox{.}
$$
Taken together, these operations leave the first and third registers
unchanged, while their action on the second and fourth defines an
oracle for the search problem. Applying Grover's
algorithm\cite{Grover96} to this oracle, we obtain the state
$|y\rangle|f^{-1}(y)\rangle$ after $O(\sqrt{N})$ invocations.

{\bf Lemma 1} \qquad To simulate the inverse oracle $S_{f^{-1}}$ with
a quantum network using oracles $S_f$ and $(S_f )^{-1}$, a total
number of $\Theta(\sqrt{N})$ invocations of $S_f$ are necessary.

\textbf{Proof} The upper bound of $O(\sqrt{N})$ is implied by the
Grover-based algorithm just discussed. Ambainis \cite{Ambainis00} has
shown that $\Omega(\sqrt{N})$ invocations of the standard oracle $S_f$
are required to invert a general permutation $f$. \qed
\vskip5pt

Given $S_f$ and $S_{f^{-1}}$, Bennett has shown how to simulate $M_f$
within classical reversible computation \cite{Bennett73}. Using a
quantum version of this construction, we can establish our main
result: \vskip5pt {\bf Lemma 2} \qquad To simulate the minimal oracle
$M_f$ with a quantum network using oracles $S_f$ and $(S_f )^{-1}$, a
total number of $\Theta(\sqrt{N})$ invocations of $S_f$ are necessary.

\textbf{Proof} Given $S_f$ and $S_{f^{-1}}$, we can simulate $M_f$ as
follows:
$$
M_f  \otimes I = ( S_{f^{-1}} )^{-1} \circ X \circ S_f \, ,
$$
where the swap gate $X$ is defined by $ X: \ket{a} \otimes \ket{b}
\rightarrow \ket{b} \otimes \ket{a}$.  From Lemma $1$, $S_{f^{-1}}$
needs $\Theta(\sqrt{N})$ invocations of $S_f$ and $(S_f
)^{-1}$. Therefore we get the upper bound of $O(\sqrt{N})$ for
simulation of $M_f$.

However this is the optimal simulation.  For suppose there is a
network which simulates $M_f$ with less than $\Omega( \sqrt{N} )$
queries. The reversed network simulates $M_{f^{-1}}$. From these two,
by our earlier results, we can construct a network that simulates
$S_{f^{-1}}$ with fewer than $\Omega ( \sqrt{N} ) $ queries, which
contradicts Lemma $1$. \qed \vskip5pt

It is worth remarking that we could equally well have carried
through our discussion using variants of $S_f$ and $P_f$,
such as the bitwise acting versions:
\begin{itemize}
\item the {\it bit string standard} oracle,
$
S^{\rm bit}_f : |{\bf x}\rangle | \bf{b} \rangle \rightarrow
|\bf{x} \rangle |\bf{b} \oplus
\bf{f(x)} \rangle $.
\item   the {\it bit string phase} oracle,
$P^{\rm bit}_f : |{\bf x}\rangle |{\bf b}\rangle
\rightarrow e^{2\pi i {\bf f(x) \cdot b } / 2 }|{\bf x}\rangle |{\bf b}
\rangle$.
\end{itemize}
Here $\bf{b} \oplus \bf{x}$ denotes the bitwise sum mod $2$ of the strings
$\bf{b}$ and $\bf{x}$, and ${\bf b \cdot x}$ their inner product mod $2$.
Again, $S^{\rm bit}_f$ and $P^{\rm bit}_f$ are equivalent:
writing
$$ {\cal F} = H \otimes H \otimes \cdots \otimes H \, ,$$
for the tensor product of $n$ Hadamard operators acting on register qubits,
we have
\begin{eqnarray*}
&(& I \otimes {\cal F} ) \circ S^{\rm bit}_f \circ
 (I \otimes {\cal F}^{-1} ) = P^{\rm bit}_f \, ,
 \\
&(& I \otimes {\cal F}^{-1} ) \circ
 P^{\rm bit}_f \circ (I \otimes {\cal F})
 = S^{\rm bit}_f \, .
\end{eqnarray*}
Note also that
$S^{\rm bit}_f = (S^{\rm bit}_f )^{-1}$,
$P^{\rm bit}_f = (P^{\rm bit}_f )^{-1}$.
Our results still apply: $S^{\rm bit}_f$ has essentially
the same relation to $M_f$ that $S_f$ does.

In summary, constructing a minimal oracle requires exponentially many
invocations of a standard oracle.  We can thus indeed definitively
exclude the possibility of efficiently solving NAGI by simulating
$M_f$ using $S_f$, which motivated our discussion. We have not,
however, been able to exclude the possibility of directly constructing
a polynomial size network defining an $M_f$ oracle for any given $1-1$
function $f$, which would lead to a polynomial time solution of
NAGI.

\noindent {\em Acknowledgments}.  We thank Charles Bennett for helpful discussions and
for drawing our attention to Refs. \cite{Bennett73}, and Richard Jozsa for helpful
comments. E. K. thanks Mike Mosca for useful discussions and Waterloo University for
hospitality.  This work was supported by EPSRC and by the European projects EQUIP,
QAIP and QUIPROCONE.

%\bibliographystyle{plain}
%\bibliography{All}

\begin{thebibliography}{10}

\bibitem{Deutsch85}
D.~Deutsch, Proc. Royal Society of London A, {\bf 400}, 97 (1985).

\bibitem{Shor94}
P.W.~Shor, SIAM J. Comp., {\bf 26}, 1484 (1997).

\bibitem{Grover96}
L.K.~Grover, Proc. 28th ACM Symp. Theor. Comp., 212 (1996).

\bibitem{BBBV97} C.~Bennett, E.~Bernstein, G.~Brassard, and
U.~Vazirani, SIAM J. Comp., {\bf 26}, 1510 (1997).

\bibitem{BBCMW98} R.~Beals, H.~Buhrman, R.~Cleve, M.~Mosca, and
R.~de~Wolf, Proc. 39th Symp. Found.  Comp. Sci., 352 (1998).

\bibitem{vanDam98}
W.~van Dam, Proc. 39th Symp. Found. Comp. Sci., 362 (1998).

\bibitem{Cleve99} R.~Cleve, in C.~Macchiavello, G.M. Palma, and
A.~Zeilinger, editors, {\em Collected Papers on Quantum Computation
and Quantum Information Theory}, (World Scientific, 1999).

\bibitem{BCWW01}
H.~Buhrman, R.~Cleve, J.~Watrous, and R.~de~Wolf, quant-ph/0102001, 2001.

\bibitem{Aaronson01}
After this work was circulated, a non-trivial lower

bound was given by S. Aaronson quant-ph/0111102.

\bibitem{BHT98} G.~Brassard, P.~H{\o}yer, and A.~Tapp, 3rd Latin
American Theor. Info. Symp., {\bf 1380}, 163 (1998).

\bibitem{BHMT00} G.~Brassard, P.~H{\o}yer, M.~Mosca, and A.~Tapp, to
appear in Quantum Computation \& Quantum Information Science, AMS
Contemporary Math Series, quant-ph/0005055, 2001.

\bibitem{Ambainis00}
A.~Ambainis, Proc. 32th ACM Symp. Theor. Comp., 636 (2000).

\bibitem{Bennett73}
C. Bennett, IBM J. Res. Dev., {\bf 17}, 525 (1973).


\end{thebibliography}

\end{multicols}
\end{document}